\documentstyle[prl,aps]{revtex}
\begin{document}
\draft
\title{
Double-exchange model: phase separation versus canted spins.
}
\author{
  M. Yu. Kagan
  $^{(a)}$,
  D. I. Khomskii
  $^{(b)}$,
 and M. V. Mostovoy $^{(b)}$}
\address{$^{(a)}$  P.L. Kapitza Institute for Physical Problems, Kosygina
Str. 2,
117334 Moscow, Russian Federation\\
$^{(b)}$ University of Groningen, Nijenborgh 4, 9747 AG, Groningen, The
Netherlands.}
\maketitle \widetext

\begin{abstract}
\begin{center}
\parbox{14cm}
{We study the competition between different possible ground
states of the double-exchange model with strong ferromagnetic
exchange interaction between itinerant electrons and local
spins. Both for classical and quantum treatment of the local
spins the homogeneous canted state is shown to be
unstable against a phase separation. The conditions for the phase
separation into the mixture of the antiferromagnetic and
ferromagnetic/canted states are given.  We also discuss another
possible realization of the phase-separated state: ferromagnetic
polarons embedded into an antiferromagnetic surrounding.  The
general picture of a percolated state, which emerges from these
considerations, is discussed and compared with results of recent
experiments on doped manganaties.}

\end{center}
\end{abstract}

\hspace{1.9cm}


\section{Introduction}

The double-exchange model describing itinerant electrons
interacting with local spins was first used to explain
ferromagnetism in metals, such as Ni and Fe \cite{Zener}.  The
revival of interest to this model was prompted by the
recent observation of the colossal magnetoresistance (CMR) in
manganites \cite{Ramirez}.  The limit of strong ferromagnetic
(FM) exchange between the conducting electrons and spins,
relevant for manganites, was first discussed by de~Gennes
\cite{pdg}.  He suggested that the competition between the
antiferromagnetic (AFM) superexchange and the double exchange
results in the canting of the AFM state (the angle $\theta$
between the spins from different sublattices becomes smaller than
$\pi$).  The canting angle grows with the concentration of charge
carriers, which explains the increase of magnetization upon
doping observed in La$_{1-x}$Ca$_x$MnO$_3$.

However, already rather long ago arguments against the canted
ground state were put forward, most notably by Nagaev \cite{Nagaev}.
In the de Gennes approach the local spins were treated
classically. Quantum corrections stabilize the AF state and the
canting appears only above certain concentration of charge
carriers \cite{Nagaev}.

A more fundamental problem is that at partial filling of the
conduction band any homogeneous ground state may be unstable
against a phase separation.  Such an instability takes place in
the Hubbard model away from half-filling \cite{Visscher} and in
the $t-J$ model \cite{Emery}.  In a wide range of parameters the
ground state of these models is phase-separated, stable
homogeneous solutions being rather an exception.  In this respect
the double-exchange model is rather similar, which can be seen
already from the tendency to a formation of magnetic polarons in
this model \cite{Nagaev,Kasuya}.  These polarons are FM droplets
surrounding charge carriers in the AF background.  If for some
reason (large effective mass, disorder) the polarons become
immobile, they can be viewed as a form of the FM-AFM phase
separation (see also \cite{Yunoki,Arovas}).

The phase separation in the double-exchange model would imply
that many experimental data on doped manganites should be
reinterpreted taking into acount the magnetic and electronic
inhomogeneity of the ground state.  In particular, the charge
transport and the metal-insulator transition should be described
in terms of percolation rather than by the properties of some
pure states.  The percolation in managanites was emphasized by
Gorkov and Kresin in Ref.~\cite{Gorkov}.  Some recent
experimental results also strongly point in this direction
\cite{Teresa,Allodi,Babushkina}.

In this paper we discuss the instability against the phase
separation in the double-exchange model. Although we mostly keep
in mind applications to CMR manganites, our treatment has a
general character. We go beyond the classical treatment of local
spins used in most numerical studies of the double-exchange
model. Our treatment has some points in common with the previous
investigations. Nevertheless, we think it is worthwhile to include
(with necessary modifications) some old results, so as to have
a full coherent picture of the behavior of this system.

Our paper is organized as follows.  After formulating the model
in Sec.~\ref{canted}, we discuss the homogeneous canted state,
taking into account quantum effects.  Several critical
concentrations separating different regimes are identified.  In
Sec.~\ref{instability} we demonstrate that for small
concentrations of charge carriers the homogeneous canted state is
unstable against a phase separation.  In Sec.~\ref{diagram} the
general phase diagram of the double-exchange model is discussed.
The Maxwell construction is used to investigate the
characteristics of the coexisting phases.  It is shown, in
particular, that depending on the parameters, there may be phase
separation not only on AFM and FM phases, but also on AFM and
canted states.  Then in Sec.~\ref{polaron}, we consider the
polaronic state and explicitely show that the energy of this
inhomogeneous state is lower than the energy of the homogeneous
canted state, even when quantum effects are taken into account.
Finally, in the concluding Sec.~\ref{conclude} we summarize the
main results and discuss the resulting picture in relation to
some recent experimental observations.

\section{Energy of homogeneous canted state}
\label{canted}

We consider the double-exchange Hamiltonian describing the
interaction of itinerant electrons with localized spins, to which
we add the Heisenberg interaction between the spins (FM
Kondo-lattice model):
\begin{equation}
H = - t \sum_{<ij> \sigma}
\left(c^{\dagger}_{i\sigma} c_{j\sigma} +
c^{\dagger}_{j\sigma} c_{i\sigma}\right)
- J_H\sum_i\vec{S}_i \vec{s_i}
+ J \sum_{<ij>} \vec{S_i}\vec{S_j}.
\label{model}
\end{equation}
Here, the first term describes the hopping of conduction
electrons, the second term is the on-site FM exchange between the
spin of a conduction electron $\vec{s}$ and the localized spin
$\vec{S}$, and the last term is the AFM exchange between the
nearest localized spins ($J_H, J > 0$). We assume the value of
localized spins to be large: $S \gg 1$. Futhermore, we consider the
case of strong Hund coupling $J_H$ and weak Heisenberg coupling
$J$:
\begin{equation}
J_H S \gg t \gg J S^2.
\label{condition1}
\end{equation}
Finally, in this paper we consider the simple cubic lattice.

The model Eq.(\ref{model}) describes the competition between the
direct antiferromagnetic exchange of the localized spins and the
double exchange via the conduction electrons, which tends to
order the spins ferromagnetically.  The simplest state that could
result from this competition, is a canted state, which has the
features of both the FM and AFM states: as in the AFM state, the
localized spins in the canted state form two sublattices,
however an angle $\theta$ between the magnetizations of the
sublattices is, in general, arbitrary, so that the total magnetic
momentum of the system is nonzero.  The canted state
interpolates between the AFM state ($\theta = \pi$) and the FM
state ($\theta = 0$).

In the classical treatment of localized spins, used
in Ref.~\cite{pdg}, the orientation of the spins is fixed:
\begin{equation}
{\vec S}_i = \left \{
\begin{array}{ll}
S {\vec n}_A & \mbox{if $i \in A$} \\
S {\vec n}_B & \mbox{if $i \in B$}
\end{array}
\right. ,
\label{cc}
\end{equation}
where $A$ and $B$ denote the two sublattices and the unit vectors
${\vec n}_A$ and ${\vec n}_B$ describe the direction of the
magnetization of the corresponding sublattice. Such a state we
shall call the classical canted (CC) state.

In another approach \cite{Nagaev}, one assumes that Eq.(\ref{cc})
only holds on sites that are not occupied by conduction
electrons.  In the quantum language, the localized spins on empty
sites have the maximal projection, $+S$, on the magnetization
vector of the corresponding sublattice.  On occupied sites,
however, the localized spin ${\vec S}$ and the spin of a
conduction electron ${\vec s}$ form a state with the maximal
total spin $S_{tot} = S + \frac{1}{2}$, because it has the lowest
energy in the large-$J_H$ limit.  The projection of the total
spin on the axis of the corresponding sublattice can have only
two values: $S_{tot}^{z} = S \pm \frac{1}{2}$ (otherwise, the
localized spin cannot remain in the state with the maximal
projection after the conduction electron leaves the site).  The
hopping of electrons between the two sublattices is
then described by a $2\times2$ matrix.  Diagonalizing this matrix
one obtains two electron bands with the corresponding hopping
amplitudes \cite{Nagaev},
\begin{equation} t_{\pm} =
t\left[ \frac{\sqrt{2S+1+S^2\cos^2\frac{\theta}{2}}}{2S+1} \pm
\frac{S \cos \frac{\theta}{2}}{(2S+1)} \right] = \frac{t}{\sqrt{2
S + 1}} e^{\pm \gamma},
\label{tpm}
\end{equation}
where $\gamma$ is defined by $\sinh \gamma = S
\cos\frac{\theta}{2} / \sqrt{2 S + 1}$. The two-band canted state
is refered below as the quantum canted (QC) state (though the
quantum fluctuations of local spins are not taken into account in
this approach).

The origin of the two hopping amplitudes $t_\pm$  can be easily
understood on the simple examples. For the AFM state ($\theta =
\pi$) the two have the same width:
$t_+ = t_- = \frac{t}{\sqrt{2 S + 1}}$. An electron, propagating
coherently in the AFM spin background, creates a state with
$S_{tot}^z = S + \frac{1}{2}$ on one sublattice and a state with
$S_{tot}^z = S - \frac{1}{2}$ on the other:
\[
\left| S_{tot}^z =
S + \frac{1}{2} \right> \rightarrow \left| S_{tot}^z = S -
\frac{1}{2} \right> \rightarrow \left| S_{tot}^z = S +
\frac{1}{2} \right> \ldots ,
\]
which closely resembles a type of motion considered by Zaanen
{\em et al.} in Ref.(\cite{zaanen}). Similarily, for the FM
ordering ($\theta = 0$) we obtain from (\ref{tpm}) $t_+ = t$ and
$t_- = \frac{t}{2S + 1}$. Here, $t_+$ corresponds to the motion
of a spin-up electron in the spin-up FM background, whereas $t_-$
describes the motion of the spin-down electron in the same
background. In the latter case, on each site the electron spin
and the local spin form the  $|S_{tot}^z =
S-{1\over2}\rangle$ state.

We denote by $n_{\pm}$ the densities of electrons occupying the
two bands, so that
\begin{equation}
n_+ + n_- = n ,
\end{equation}
where $n$ is the total electron density.  As we shall shortly
see, under condition (\ref{condition1}) the transitions between
the AFM, canted, and FM states take place at $n \ll 1$.  In this
case all occupied electron states lie close to the bottom of each
of the two bands, and the momentum dependence of the electron
energy can be approximated by:
\[
\varepsilon_{\pm}\left(\vec{p}\,\right) = - z t_{\pm} +
\frac{p^2}{2 m_{\pm}} ,
\]
where $\frac{1}{2m_{\pm}}=t_{\pm}$ (the lattice constant is put
to $1$) and $z = 6$ is the number of nearest neighbors.  The
densities $n_{\pm}$ are related to the corresponding Fermi
momenta $p_{F\pm}$ by
\begin{equation}
n_{\pm}=\frac{p_{F{\pm}}^3}{6\pi^2} ,
\label{xpm}
\end{equation}
and the chemical potential of conducting electrons is:
\begin{equation}
\mu = -zt_+ + \frac{p_{F+}^2}{2m_+} = -zt_- +
\frac{p_{F-}^2}{2m_-} .
\label{mu}
\end{equation}
The second part the Eq.(\ref{mu}) only holds if both electron
bands are occupied.

The energy (per site) can then be written as follows:
\begin{equation}
E =  \frac{z}{2}J S^2 \cos \theta
- zt_+ n_+ - zt_- n_-
+ a t_+ n_+^{5/3} + a t_- n_-^{5/3},
\label{E}
\end{equation}
where the first term is the energy of the AFM exchange between
localized spins and other terms represent the energy of
conducting electrons ($a = \frac{3}{5} (6 \pi^2)^{2/3}$).

The large energy of the on-site FM exchange, $- J_H \frac{S}{2}
n$, is substracted from Eq.(\ref{E}), as it only renormalizes the
chemical potental. For $n \ll 1$, the fourth and fifth terms in
Eq.(\ref{E}) (the kinetic energy of electrons counted from the
bottom of the corresponding band) are small compared to,
respectively, the second and the third terms. Therefore, in this
section the last two terms will be neglected (their role is
considered in Sec. \ref{instability}).

We first consider the case when the canting angle $\theta$ is
very close to $\pi$ and $\gamma \ll 1$, so that, according to
Eq.(\ref{tpm}), the difference between $t_+$ and $t_-$ is
relatively small and the two electron bands are filled almost
equally: $\frac{(n_{+} - n_{-})}{n_{+}} \ll 1$. Then, using
Eqs.(\ref{tpm})-(\ref{mu}), the energy of the QC state is
approximately:
\begin{equation} E_{QC} = -
\frac{tzn}{\sqrt{2S+1}} - \frac{z}{2} JS^2 + \frac{z}{3} J (2 S +
1) \left[ - \frac{(n - n_1)}{n_1} \gamma^2 + \frac{z^6 t^4}{32
\pi^8 (2 S + 1) J^4 } \gamma^4 \right] ,
\label{Ecant}
\end{equation}
where
\begin{equation}
n_1=\frac{8
\pi^4}{3} \left(\frac{J(2 S + 1)^{3/2}}{z t} \right)^3 .
\end{equation}
The optimal value of $\gamma$ can be found from
the condition {\Large $\frac{dE}{d\gamma^2}$} $=0$ and is given
by:
\begin{equation} \gamma^2 = \left\{ \begin{array}{ll} 0 &
\mbox{for $n < n_1$} \\ &  \\ \normalsize{\frac{6 \pi^4 J (2 S +
1)^{3/2}}{z^3 t} (n - n_1)} & \mbox{for $n < n_1$} \end{array}
\right. \label{gamma}
\end{equation}
(the last expression is only valid for $(n - n_1) / n_1 \ll 1$).

While in the classical treatment of de Gennes the canting of
sublattices takes place at arbitrarily small doping \cite{pdg},
in the quantum treatment the system stays antiferromagnetic
up to the critical concentration $n_1$ \cite{Nagaev}.  This
happens because in the two-band picture electrons can propagate
even in the colinear AFM state via the process described
above, while in the classical treatment they are localized. The
two bands have the same width, $\frac{2 z t}{\sqrt{2S+1}}$, and
are equally occupied, so that the energy of the AFM state is:
\begin{equation}
E_{AFM} =
- \frac{tzn}{\sqrt{2S+1}} - \frac{z}{2}JS^2.
\label{EAFM}
\end{equation}

For $n > n_1$, the energy of the two-band QC state
becomes lower than the energy of the AFM state:
\begin{equation}
E_{QC}= E_{AFM}
- \frac{3}{8} \frac{z t^2}{J (2 S + 1)^2}
\left(n - n_1\right)^2 .
\label{EQCxc1}
\end{equation}

As $\gamma$ grows with increasing $n$, so does the difference
between $t_+$ and $t_-$.  As a result, above certain
concentration $n_2$ the second part of Eq.(\ref{mu}) can no
longer be satisfied.  Thus, for $n > n_2$ only the ``$+$'' band
is filled, while the ``$-$'' band is empty ($n_+ = n$ and $n_- =
0$).

The energy of the one-band QC state is given by
\begin{equation}
E_{QC} = - z t_+ n -
\frac{z}{2} J S^2 +
z J (2 S + 1) \sinh^2 \gamma .
\end{equation}
An optimal value of $\gamma$ for $n > n_2$ satisfies
\begin{equation}
e^{-\gamma} \sinh{2 \gamma} = \frac{n t}{J (2 S + 1)^{3/2}} .
\label{eqgamma}
\end{equation}
The concentration $n_2$, at which
the second band disappears, is found using Eq.(\ref{eqgamma}) and
\[
z(t_+ -t_-) = \frac{p_F^2}{2m_+} =t_+n_2^{2/3}
\]
[cf. Eq.(\ref{mu})] to be
\begin{equation}
n_2 \approx \frac{27}{2} n_1 .
\end{equation}

When $\gamma$ becomes much larger than $1$, the hopping amplitde
$t_+ \approx t \cos \frac{\theta}{2}$, i.e., it coincides with
the classical expression used by de Gennes.  In this case we
recover the CC state with the energy
\begin{equation}
E_{CC} = - z t n
\cos\frac{\theta}{2} + \frac{z}{2} JS^2 \cos \theta.
\label{ECC}
\end{equation}
The optimal canting angle is given by
$\cos\frac{\theta}{2}=\frac{tn}{2JS^2}$, so that the
optimal energy of the CC state (for $S \gg 1$) is
\begin{equation}
E_{CC} = - \frac{zt^2n^2}{4JS^2} - \frac{z}{2}JS^2 .
\label{ECCO}
\end{equation}
The cross-over from the one-band QC state to the CC state occurs
at
\begin{equation}
n \sim n_3 =  \frac{8 J S^{3/2}}{t} .
\end{equation}

Finally, for
\begin{equation}
n > n_4 = \frac{2 J S^2}{t},
\label{21prime}
\end{equation}
the angle
$\theta = 0$, {\em i.e.}, the canted state transforms into the FM
state with the energy:
\begin{equation}
E_{FM} = - ztn + \frac{z}{2}JS^2.
\label{EFM}
\end{equation}

Summarizing the results of these section, we can say that on the
class of the homogeneous trial ground states with two
sublattices, the AFM state is the most favorable for $n < n_1$.
For $n_1<n<n_2$ the two-band QC state has lower energy.  For
$n>n_2$ only one of the bands is filled.  For $n{\
\lower-1.2pt\vbox{\hbox{\rlap{$ >
$}\lower5pt\vbox{\hbox{$\sim$}}}}\ } n_3$ the one-band QC state
coincides with the CC state.  Finally, for $n>n_4$ the ground
state is ferromagnetic.

\section{Instability of canted state against phase separation}
\label{instability}

In this section we study the stability of the canted state
against a phase separation by calculating the
electronic compressibility,
\[
\kappa = \frac{d^2E}{dn^2} .
\]

It is easy to see that under the condition (\ref{condition1}) and
for small $n$ the compressibility is negative. Thus, for the
two-band QC state at $n{\ \lower-1.2pt\vbox{\hbox{\rlap{$ >
$}\lower5pt\vbox{\hbox{$\sim$}}}}\ } n_1$
\begin{equation}
\kappa = - \frac{3}{4} \frac{z t^2}{J (2 S + 1)^2},
\end{equation}
[cf. Eq.(\ref{EQCxc1})], while in the classical canted regime
[see Eq.(\ref{ECC})]
\begin{equation}
\kappa = - \frac{z t^2}{2 J S^2}.
\end{equation}

Thus we see, that the homogeneous canted (both classical and
quantum) state is unstable for small $n$. This instability is
similar to the instability against a phase separation in the
Hubbard model \cite{Visscher} and $t$-$J$ model \cite{Emery}.

At sufficiently large $n$ the kinetic electron energy makes the
compressibility of the canted state positive. For arbitrary
band-filling Eq.(\ref{ECC}) for the energy of the CC state has
to be substituted by
\begin{equation}
E_{CC}(n) = - t \cos \frac{\theta}{2} f(n) - \frac{{\bar J}}{2}
\cos \theta,
\label{ECCn}
\end{equation}
where $f(n)$ is the kinetic energy of electrons with
the hopping amplitude $t = 1$ (which depends only on the
electron density $n$), and ${\bar J} = z J S^2$. Then
the canting angle is given by
\begin{equation}
\cos \frac{\theta}{2} = \frac{t f(n)}{2{\bar J}}
\end{equation}
and the energy of the CC state is
\begin{equation}
E_{CC}(n) = - \frac{t^2}{4 {\bar J}} f(n)^2
- \frac{\bar J}{2}.
\label{E_CC(n)}
\end{equation}
For small band-filling $f(n) \approx - z n$ and the last equation
agrees with Eq.(\ref{ECCO}).  While for small $n$ the
compressibility is negative, the increase of the electron density
results in the increase of the Fermi-pressure of the electron
gas, which stabilizes the canted state.  At half-filling the
electron energy, $f(n)$, reaches minimum and, according to
Eq.(\ref{E_CC(n)}), so does the energy of the CC state.  Thus,
at least close to half-filling $\kappa > 0$.  For $D=3$, the
compressibility of the CC state is positive for $n > \sim 0.16$.

\section{Phase diagram}
\label{diagram}

In this section we discuss the phase diagram of the
double-exchange model. We treat spins classically
($S \rightarrow \infty$), keeping, however, ${\bar J} = z J S^2$
finite (the numerical calculations in this section were performed
for $D=3$, but the behavior in other dimensions is qualitatively
the same). In Fig.~1 we plot the energy of a homogeneous
state, $E(n)$, which is the CC state, for $n < n_4$, and the FM
state for $n > n_4$. The energy of the CC state is given by
Eq.(\ref{E_CC(n)}), while the energy of the FM state for
arbitrary density is
\[
E_{FM}(n) = t f(n) + \frac{\bar J}{2}.
\]

For small $n$, the CC state separates on the pure AFM phase and
a second phase that contains all the electrons. The density
of electrons in the second phase, $n_{\ast}$, is defined by the
Maxwell construction:
\begin{equation}
\frac{dE(n_{\ast})}{dn_{\ast}} = \frac{E(n_{\ast})
+ \frac{z}{2} {\bar J}}{n_{\ast}}.
\label{Maxwell}
\end{equation}

For $n_{\ast} > n_4$, the second phase is ferromagnetic (see
Fig.~1a), while for $n_{\ast} < n_4$ it is the stable canted
state (see Fig.~1b). Which of the two situations is realized
depends on the ratio of ${\bar J} \over t$. For the small ratio
(${{\bar J} \over t} < \sim 0.425$ in 3D), the phase
separation goes into the pure AFM and FM states, for larger
${\bar J} \over t$, the ground state is the mixture of the AFM
and stable canted states.  This is illustrated in Fig.~2,
where we plot the phase diagram of the system.  The thick line
shows the dependence of $n_{\ast}$ on ${\bar J} \over t$ and
separates stable states from unstable states.  In the
ferromagnetic region of the diagram $n_{\ast}$ grows with ${\bar
J} \over t$, while in the canted region $n_{\ast}\approx0.291$
independent of ${\bar J} \over t$ (both sides of
Eq.(\ref{Maxwell}) with $E = E_{CC}$ can be divided by $t^2 \over
{\bar J}$, after which the equation becomes independent of $t$
and ${\bar J}$).  In the unstable FM region and a part of the
unstable canted region the compressibility is positive, so that
the corresponding states are locally stable.  They are, however,
globally unstable against the phase separation, as follows form
the Maxwell construction.

In the $J_H \gg t$ and $S \gg 1$ limit, we consider in this
section, the phase diagram of the double-exchange model is
symmetric around $n = \frac12$. This can be seen by applying to
the Hamiltonian (\ref{model}) the particle-hole transformation
\[
c_i \rightarrow
\left \{
\begin{array}{ll}
+ h_i^{\dagger} & \mbox{for $i \in A$}\\ \\
- h_i^{\dagger} & \mbox{for $i \in B$}
\end{array}
\right. ,
\]
where $c_i$ annihilates electron on site $i$ with spin parallel
to the local spin on the same site, while $h_i$ is a
corresponding hole operator. This transformation leaves the
kinetic energy term (first term in (\ref{model})) unchanged, but
it reverses the sign of the Hund's coupling (the second term in
(\ref{model})). The fact that holes are coupled to local spins
antiferromagnetically, rather than ferromagnetically, does not,
however, make any difference: the important thing is that the
orientation of the spin of the hole is uniquely determined by the
orientation of the local spin. Thus, the double-exchange system
with $n$ holes per site (e.g., with electron density $1-n$) has the
same properties as the system with $n$ electrons per site.

Experimentally, however, there is a marked difference in the
behavior of the hole-doped manganties (La$_{1-x}$M$_x$MnO$_3$, $M
=$ Sr, Ca, $x < 0.5$) and the electron-doped (overdoped)
manganites ($x > 0.5$, the electron density related to doping
by $n = 1 - x$).  While the properties of the hope-doped systems
qualitatively resemble those described above, the electron-doped
manganites (with predominantly Mn$^{4+}$) behave quite
differently: they are mostly insulating, often with charge
ordering in the form of stripes \cite{Mori} and sometimes with
anisotropic magnetic structures \cite{Akimoto}.  The reason for
the assymetry of the phase diagram is not completely clear.  One
of the factors, which is not taken into account in the
double-exchange model studied above, but which may play an
important role, is the orbital degeneracy of the electrons bands
\cite{vdBrink}.  Note, however, that the tendency to a phase
separation discussed in this paper is also present for the
degenerate bands \cite{vdBrink}.

\section{Polaron state}
\label{polaron}

There exist many different realizations of an inhomogeneous
phase-separated state.  For the model (\ref{model}) the most
favorable one is the complete phase separation (considered in the
previous section), in which case one part of the sample contains
all the charge carriers and is ferromagnetic (or strongly
canted), while the other part is an undoped antiferromagnet.
Such a phase separation would, however, be counteracted by the
Coulomb interaction (not included in the Hamiltonian
(\ref{model})).  The Coulomb interaction may suppress the phase
separation completely, or it could favor the formation of the FM
droplets (spin-bags) immersed into the AFM matrix.  More
complicated structures, e.g., stripes \cite{Zaanen} are also
possible.  In general, the phase separation in the
double-exchange model in the presence of Coulomb interactions
seems to be rather similar to that in the t-J model considered in
the context of the high-T$_c$ superconductors
\cite{Emery,Zaanen}.  In the later case, the instability towards
the phase separation remains even with the Coulomb interaction
taken into account.

In this section we consider a simple type of the phase separated
state: the AFM matrix with the FM ``droplets'', each containing
one electron.  In other words, we assume that electrons
propagating in the AFM background form ferromagnetic polarons
\cite{Nagaev,Kasuya}.  Such a state, with only one electron per
droplet and with the droplets well separated from each other will
help to minimize the Coulomb energy contribution mentioned above.
We will demonstrate that the energy of this state is, indeed,
lower than that of the homogeneous canted state.

We calculate now the energy of the system for small doping, $n
\ll 1$, assuming that each electron is localized inside a
FM sphere of radius $R$. The energy of such a state
can is given by
\begin{equation}
E_P = -
- tn \left( z - \frac{\pi^2}{R^2} \right) + n
\frac{\bar J}{2} \frac{4}{3} \pi R^3 - \frac{\bar J}{2}
\left( 1- n \frac{4}{3}\pi R^3 \right).
\label{Epol}
\end{equation}
The first term in (\ref{Epol}) is the energy of the lowest state
of an electron in the sphere, while the second and
the third terms give, respectively, the exchange energy of
localized spins in the ferromagnetic regions and in the remaining
antiferromagnetic matrix.  The ferromagnetic polarons are
assumed to have sharp boundaries and the surface energy is
neglected.  Furthermore, we assume that the polarons do not
overlap, in which case the change of the energy of the system due
to doping is simply proportional to $n$.

The optimal polaron radius, found from the condition
{\Large$\frac{dE_{P}}{dR}$}$=0$, is given by
\begin{equation}
  R = \left( \frac{\pi t}{2{\bar J}}\right)^{1/5}
\end{equation}
and the energy of the polaron state is:
\begin{equation}
  E_{P}= - \frac{\bar J}{2} - ztn
  + \frac{5\pi}{3} n (\pi t)^{3/5} (2 {\bar J})^{2/5}.
\end{equation}

It is easy to see that the energy of such a polaron state is
lower than the energy of both the classical and the quantum
homogeneous canted states.  Moreover, at $n = n_4$ the polaron
state is also more favorable than the FM state.  Only at higher
concentration,
\[
\frac{3}{10\pi} \left(\frac{2 {\bar J}} {\pi t}\right)^{3/5},
\]
the transition to the saturated FM state occurs.  At this
concentration polarons begin to overlap and for small ${\bar J}
\over t$ this concentration up to a numerical factor coincides
with $n_{\ast}$ discussed in the previous section.

Ferromagnetic polarons can be also formed in materials with
layered electronic and magnetic structure, such as manganites.
In LaMnO$_3$ the layers are formed by planes of ferromagnetically
coupled localized spins, while the exchange between the spins
from two neighboring layers is antiferromagnetic.  Also, the
interlayer and intralyer hopping amplitudes, $t_{\parallel}$ and
$t_{\perp}$, are different.  Such a layered magnetic and
electronic structure is a result of the orbital ordering that
takes place in this material \cite{Khomskii}.  The polaron is
elongated (``cucumber'') for $t_{\parallel} > t_{\perp}$ and
compressed (``pancake'') for $t_{\parallel} < t_{\perp}$. A
detailed discussion of the FM polaron in the layered system will
be given elsewhere.

\section{Conclusions}
\label{conclude}

In conclusion, we have shown that the tendency to a phase
separation and a formation of a spatially inhomogeneous state,
found for several models of strongly correlated electrons, is
also an inherent property of the double-exchange model.  For
large Hund's coupling, $J_H \gg t \gg JS^2$, and small doping the
homogeneous canted state has negative compressibility.  This is
related to the instability of the canted state against the phase
separation into pure antiferromagnetic and ferromagnetic phases.
The latter phase contains all charge carriers.  For $zJS^2 \sim
t$ the compressibility of the canted state is negative only at
small concentrations of charge carriers.  In that case, it can
separate into the pure antiferromagnetic phase and the canted
state with a large concentration of charge carriers and a large
canting angle.

The spatially inhomogeneous state may occur in a variety of
different forms, e.g., a random mixture of AFM and FM phases,
magnetic polarons (``fog'' of small FM droplets), or some regular
structure, such as stripes. Which of this situations is realized
would depend on specific conditions and may be different in
different parts of the phase diagram.  However, the tendency to
form an inhomogeneous state seems to be a generic property of the
double-exchange model.  The specific features of manganaties
(layered magnetic structure, orbital degeneracy), though
definitely important for determining a particular form of the
ground state, nevertheless seem to preserve this tendency.

Several theoretical problems still remain to be clarified. The
first one, already mentioned above, is the role of the Coulomb
interaction in determining a detailed nature of the inhomogeneous
state (the average size of polarons and their spatial ordering
(cf.  experimental results \cite{Hennion})).  Another question is
the detailed spin structure both inside and outside polarons.  In
the treatment above we made a simplifying assumption that the inner
part is the saturated ferromagnet, while the outer part is the
pure antiferromagnet without canting.  However, in general, this
need not to be the case: both these phases may have certain
degree of canting.  In particular, since the spectrum of spin
excitations in AF matrix is gapless, the distortion of the
perfect antiferromagnetic order should decay slowly with the
distance from a polaron \cite{pdg}.  All these questions
definitely require further study, but at this point we can, in
any case, say that the homogeneous canted state seems to be
nearly always unstable, and inhomogeneous structures of some kind
have to be realized in low-doped double exchange systems,
including CMR manganites.

Turning to an experimental situation in manganites, we can state that
there exist already many indications that the phase separation is
indeed present there. The formation of stripe phases in overdoped
La$_{1-x}$Ca$_x$MnO$_3$ ($x > 0.5$) \cite{Mori} can be viewd as a
manifestation of this tendency.  However, much more indications
of this are now observed in the most interesting region of $x <
0.5$ and, in particular, in the CMR regime.  Already the old data
\cite{Wollan} on a coexistence of both the FM and AFM phases in
the magnetic neutron scattering, which are sometimes treated in a
picture of homogeneous canted state, are actually much better
interpreted as a mixture of FM and AFM microregions
\cite{Nagaev}.  Recent neutron scattering results \cite{Hennion}
confirm this picture and, moreover, show the presence of spatial
liquid-like correlations of the FM droplets (in the system
studied in \cite{Hennion} these were not purely FM droplets, but
rather the microregions with the large canting angle).  Direct
confirmation of the existence of the inhomogeneous state comes
from the NMR measurements \cite{Allodi}, which show the presence
of two different hyperfine fields corresponding to FM and AFM
regions.  And finally, quite recently the presence of an
inhomogeneous state with a random mixture of phases with
different electronic properties, was directly visualized by the
STM study of La$_{0.7}$Ca$_{0.3}$MnO$_3$ \cite{Faeth}: the
metallic and insulating regions with an average scale of
$10-50$nm were directly seen in this investigation.

The natural consequence of the phase separation is that the main
properties of CMR manganites, including transport properties,
should be treated in the percolation picture \cite{Gorkov}.  This
concept explains quite naturally the critical doping $x_c \sim
0.16$, above which the ferromagnetic metallic state appears in
La$_{1-x}$M$^{2+}_x$MnO$_3$ \cite{Gorkov}.  An experimental
support for the percolation picture comes from the recent study
\cite{Babushkina}, which shows that $I-V$ characteristics of
La$_{1-x}$Pr$_x$Ca$_{0.3}$MnO$_3$ are strongly nonlinear even at
small voltages: such behavior is expected for the percolated
state, for which the electric field would modify the distribution
and shape of metallic clusters and enhance conductivity.

Thus, both from the theoretical treatment and from many experimental
results it follows that, most probably, the electronic and magnetic
state of doped manganites is intrinsically inhomogeneous, leading
to the percolated state.  This should be definitely the case for
small doping (for which the random potential of impurities should
be also taken into account); but it is also quite probable that
such is the nature of even optimally doped manganites ($x \sim
0.3$) with CMR.  This gradually emerging picture, promisses to
give a new explanation of the properties of these fascinating
materials.

Authors are greatful to E.L.  Nagaev, L.P.  Gor'kov, M.  Hennion,
N.A.  Babushkina, K.I.  Kugel, R.  Ibarra, J.M.  Coey, G.A.
Sawatzky, J.  Aarts, J.  Mydosh, P.  Guinea and many other
collegues for useful discussions and for information of the new
experimental results prior to their publication.  This work was
supported by NWO grant and President Eltsin grant
\#97-26-15H/363, by the Dutch Foundation for Fundamental Studies
of Matter (FOM) grant, and by the Eropean Network OXSEN.

\newpage
\begin{center}
\Large
Figure Captions
\end{center}

\vspace{1cm}

Fig 1.  The energy of a homogeneous state $E$ as a function of
electron density $n$ (thick line) calculated for $t=1$ and two
different values of ${\bar J}$: $\sim0.344$, corresponding to
$n_4 = 0.2$, (a) and $\sim.484$, corresponding to $n_4 = 0.4$.
The Maxwell construction (thin line) gives the density $n_{\ast}$
of the phase containing all the electrons in the phase-separated
state.  For $n_4 < n_{\ast}$, the phase-separated state is the
mixture of the AFM and FM states (a), while for $n_4 > n_{\ast}$
it is the mixture of the AFM and CC states (b).

\vspace{12pt}

Fig 2. The phase diagram of the double-exchange model.  The
states to the left of the thick line are unstable against the
phase separation.  The thick line shows the dependence of the
density $n_{\ast}$ of the electron-reach phase on ${\bar J}$.
The thin line separates the canted state from the ferromagnetic
state. The regions of the CC state with positive and negative
compressibility are separated by the dotted line.

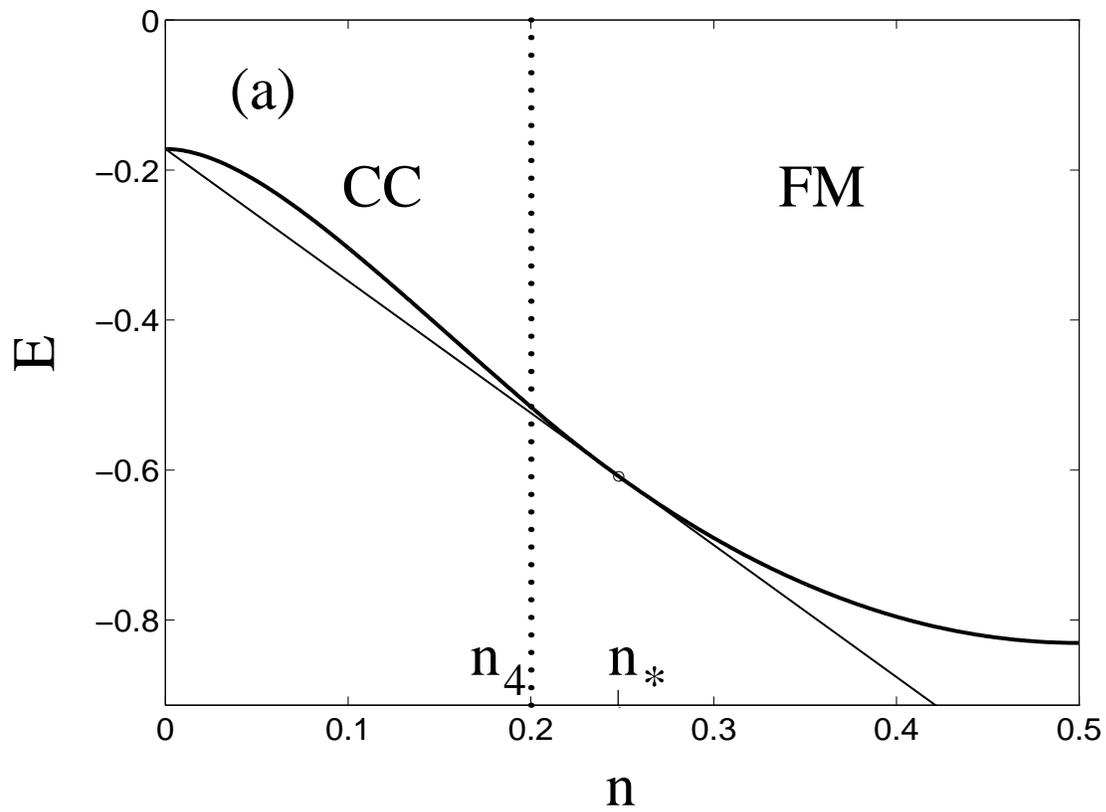
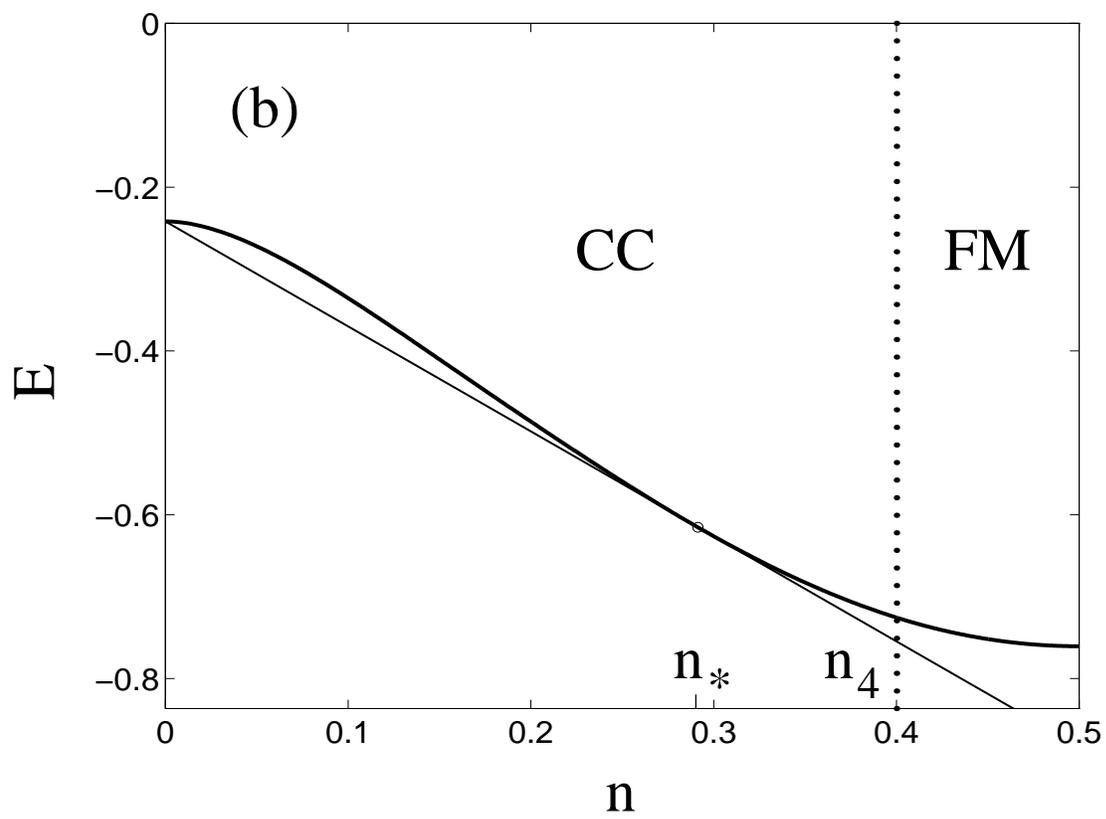
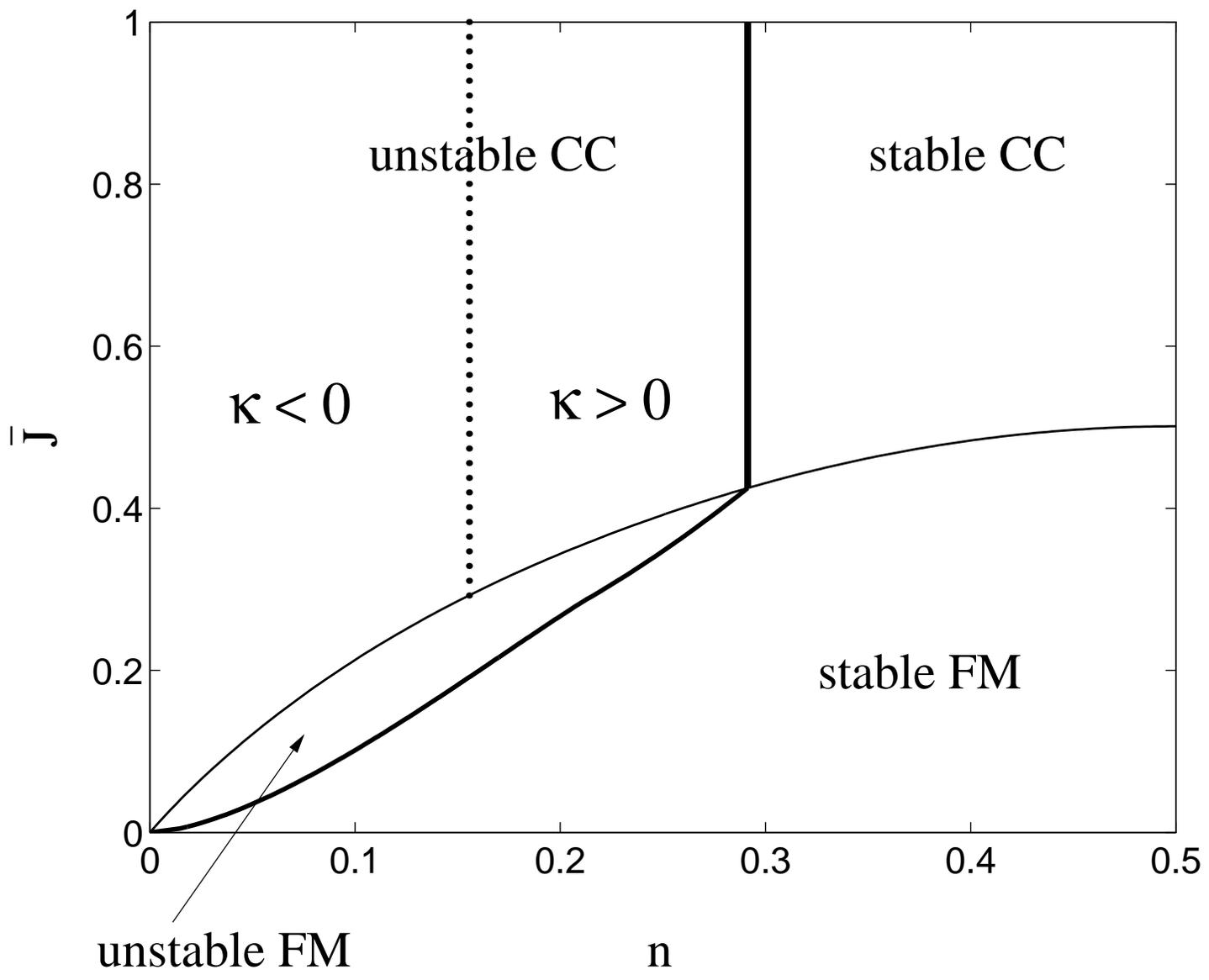
\end{document}